\documentclass[article,twocolumn]{aastex61}

\usepackage{amsmath,amssymb,amsfonts}
\usepackage{natbib}
\usepackage{graphicx}% Include figure files
\usepackage{dcolumn}% Align table columns on decimal point
\usepackage{bm}% bold math\linenumbers
\usepackage{xspace}
\usepackage{color}
\usepackage{hyperref}
\usepackage{url}
\usepackage{lineno}

\newcommand{\LIGO}{LIGO\xspace}
\newcommand{\Fermi}{\emph{Fermi}\xspace}
\newcommand{\PyCBC}{{\tt PyCBC}\xspace}
\newcommand{\GstLAL}{{\tt GstLAL}\xspace}

\newcommand{\SNR}{S/N\xspace}
\newcommand{\FARGW}{FAR$_{\rm GW}$\xspace}
\newcommand{\FARGBM}{FAR$_{\rm GBM}$\xspace}
\newcommand{\FAP}{$p$-value}
\newcommand{\FAPs}{$p$-values}
\newcommand{\msun}{M$_\odot$}

\newcounter{notes}

\begin{document}

%opening
\title{A \Fermi Gamma-ray Burst Monitor Search for Electromagnetic Signals Coincident with Gravitational-Wave Candidates in Advanced \LIGO's First Observing Run}

\AuthorCollaborationLimit=3000
%
%
%
% Contributing authors (ranked)
\author{E.~Burns}
\affiliation{NASA Goddard Space Flight Center, Greenbelt, MD 20771, USA}
\author{A.~Goldstein}
\affiliation{Science and Technology Institute, Universities Space Research Association, Huntsville, AL 35805, USA}
\author{C.~M.~Hui}
\affiliation{NASA Marshall Space Flight Center, Huntsville, AL 35812, USA}
%
%
%
% GBM-LIGO authors (alphabetical)
\author{L.~Blackburn}
\affiliation{Harvard-Smithsonian Center for Astrophysics, 60 Garden St, Cambridge, MA 02138, USA}
\affiliation{LIGO, Massachusetts Institute of Technology, Cambridge, MA 02139, USA}
\author{M.~S.~Briggs}
\affiliation{Department of Space Science, University of Alabama in Huntsville, Huntsville, AL 35899, USA}
\affiliation{Center for Space Plasma and Aeronomic Research, University of Alabama in Huntsville, Huntsville, AL 35899, USA}
\author{V.~Connaughton}
\affiliation{Science and Technology Institute, Universities Space Research Association, Huntsville, AL 35805, USA}
\author{R.~Hamburg}
\affiliation{Department of Space Science, University of Alabama in Huntsville, Huntsville, AL 35899, USA}
\affiliation{Center for Space Plasma and Aeronomic Research, University of Alabama in Huntsville, Huntsville, AL 35899, USA}
\author{D.~Kocevski}
\affiliation{NASA Marshall Space Flight Center, Huntsville, AL 35812, USA}
\author{P.~Veres}
\affiliation{Center for Space Plasma and Aeronomic Research, University of Alabama in Huntsville, Huntsville, AL 35899, USA}
\author{C.~A.~Wilson-Hodge}
\affiliation{NASA Marshall Space Flight Center, Huntsville, AL 35812, USA}
%
%
%
% Team authors (alphabetical)
\author{E.~Bissaldi}
\affiliation{Politecnico di Bari, Via Edoardo Orabona, 4, I-70126 Bari BA, Italy}
\affiliation{Istituto Nazionale di Fisica Nucleare, Sezione di Bari, I-70126 Bari, Italy}
\author{W.~H.~Cleveland}
\affiliation{Science and Technology Institute, Universities Space Research Association, Huntsville, AL 35805, USA}
\author{M.~M.~Giles}
\affiliation{Jacobs Space Exploration Group, Huntsville, AL 35806}
\author{B.~Mailyan}
\affiliation{Center for Space Plasma and Aeronomic Research, University of Alabama in Huntsville, Huntsville, AL 35899, USA}
\author{C.~A.~Meegan}
\affiliation{Center for Space Plasma and Aeronomic Research, University of Alabama in Huntsville, Huntsville, AL 35899, USA}
\author{W.~A.~Paciesas}
\affiliation{Science and Technology Institute, Universities Space Research Association, Huntsville, AL 35805, USA}
\author{S.~Poolakkil}
\affiliation{Department of Space Science, University of Alabama in Huntsville, Huntsville, AL 35899, USA}
\affiliation{Center for Space Plasma and Aeronomic Research, University of Alabama in Huntsville, Huntsville, AL 35899, USA}
\author{R.~D.~Preece}
\affiliation{Department of Space Science, University of Alabama in Huntsville, Huntsville, AL 35899, USA}
\author{J.~L.~Racusin}
\affiliation{NASA Goddard Space Flight Center, Greenbelt, MD 20771, USA}
\author{O.~J.~Roberts}
\affiliation{Science and Technology Institute, Universities Space Research Association, Huntsville, AL 35805, USA}
\author{A.~von~Kienlin}
\affiliation{Max-Planck-Institut f\"{u}r extraterrestrische Physik, Giessenbachstrasse 1, 85748 Garching, Germany}
\collaboration{\Fermi Gamma-Ray Burst Monitor}

\medskip
%\AuthorCollaborationLimit=3000  % AAS default is too small!

\author{B.~P.~Abbott}
\affiliation{LIGO, California Institute of Technology, Pasadena, CA 91125, USA}
\author{R.~Abbott}
\affiliation{LIGO, California Institute of Technology, Pasadena, CA 91125, USA}
\author{T.~D.~Abbott}
\affiliation{Louisiana State University, Baton Rouge, LA 70803, USA}
\author{F.~Acernese}
\affiliation{Universit\`a di Salerno, Fisciano, I-84084 Salerno, Italy}
\affiliation{INFN, Sezione di Napoli, Complesso Universitario di Monte S.Angelo, I-80126 Napoli, Italy}
\author{K.~Ackley}
\affiliation{OzGrav, School of Physics \& Astronomy, Monash University, Clayton 3800, Victoria, Australia}
\author{C.~Adams}
\affiliation{LIGO Livingston Observatory, Livingston, LA 70754, USA}
\author{T.~Adams}
\affiliation{Laboratoire d'Annecy de Physique des Particules (LAPP), Univ. Grenoble Alpes, Universit\'e Savoie Mont Blanc, CNRS/IN2P3, F-74941 Annecy, France}
\author{P.~Addesso}
\affiliation{University of Sannio at Benevento, I-82100 Benevento, Italy and INFN, Sezione di Napoli, I-80100 Napoli, Italy}
\author{R.~X.~Adhikari}
\affiliation{LIGO, California Institute of Technology, Pasadena, CA 91125, USA}
\author{V.~B.~Adya}
\affiliation{Max Planck Institute for Gravitational Physics (Albert Einstein Institute), D-30167 Hannover, Germany}
\affiliation{Leibniz Universit\"at Hannover, D-30167 Hannover, Germany}
% ...

\collaboration{The LIGO Scientific Collaboration and the Virgo Collaboration}

% \linenumbers

\begin{abstract}
We present a search for prompt gamma-ray counterparts to compact binary coalescence gravitational wave (GW) candidates from Advanced LIGO's first observing run (O1).  As demonstrated by the multimessenger observations of GW170817/GRB 170817A, electromagnetic and GW observations provide complementary information about the astrophysical source and, in the case of weaker candidates, may strengthen the case for an astrophysical origin. Here we investigate low-significance GW candidates from the O1 compact-binary coalescence searches using the \Fermi Gamma-Ray Burst Monitor (GBM), leveraging its all-sky and broad energy coverage.  
Candidates are ranked and compared to background to measure the significance.  Those with false alarm rates of less than $10^{-5}$ Hz (about one per day, yielding a total of 81 candidates) are used as the search sample for gamma-ray follow-up.
No GW candidates were found to be coincident with gamma-ray transients independently identified by blind searches of the GBM data.
In addition, GW candidate event times were followed up by a separate targeted search of GBM data. Among the resulting GBM events, the two with lowest false alarm rates were the gamma-ray transient GW150914-GBM presented in \cite{GW150914_GBM} and a solar flare in chance coincidence with a GW candidate.  
%The analyses reported here will be performed in low-latency and over a wider scope of candidates during future gravitational-wave observing campaigns to rapidly identify EM counterparts to GW sources.
\end{abstract}

% \maketitle

\section{Introduction}
The first observing run (O1) of the Advanced LIGO detectors~\citep{aLIGO} marked the dawn of gravitational wave (GW) astronomy with the ground-breaking discovery of merging black holes (BHs)~\citep{GW150914,GW151226} and \cite{O1_BBH}. 
The second observing run (O2) continued unveiling the population of binary black holes~\citep{GW170104, GW170608, GW170814}, saw the addition of the Virgo observatory to the detector network~\citep{GW170814}, and culminated in the spectacular multimessenger observations of a binary neutron star merger, summarized in~\citet{GW170817_BNS,GW170817_MMA}.
%As GW detectors improve in sensitivity and embark on longer observing campaigns, a world-wide effort to discover the first unambiguous electromagnetic (EM) counterpart to a GW source is underway~\citep{GW150914_EM}.  
Simultaneously observing the same astrophysical event in both gravitational and electromagnetic radiation will continue to uniquely enrich our understanding of sources. 
Because GWs negligibly interact with matter they are directly encoded with information about the central engines of the most violent, dynamic processes in the universe. 
Electromagnetic (EM) waves, on the other hand, are tightly coupled to matter thus providing information about the material and surrounding environment being affected by the central engine~\citep{Metzger:2012}. 
%Joint GW and EM searches can also increase confidence in putative detections in one band or the other.
%Therefore, joint observations of GW and EM radiation are greater than the sum of their parts.

Several astrophysical transient phenomena are thought to produce GW and EM emission strong enough to be detected by current or proposed observatories including soft gamma-ray repeaters, rapidly rotating core collapse supernovae, binary neutron star mergers and gamma-ray bursts. 
Here we focus on short gamma-ray bursts (SGRBs), now directly confirmed to arise from the mergers of compact stellar remnants to which ground-based GW detectors are most sensitive~\citep{GW170817_GRB}.
This paper is limited in scope to analysis of times during O1, and focuses on the follow-up searches of data from the \emph{Fermi} Gamma-ray Burst Monitor (GBM) near in time to the GW search candidates.
Results for the search in GW data for known GRBs which occurred during O1 can be found in~\cite{O1_GRB}.  
Despite the consensus view that there would be no bright EM emission associated with stellar-mass BBH mergers, comprehensive observing campaigns with EM observatories were carried out.  For example, twenty five participating teams of observers received and responded to notifications of the GW150914 detection, with follow-up observations taken from the radio to gamma-ray bands~\citep{GW150914_EM}. In the follow-up analysis of the time around the first BBH merger observation, GW150914, the \Fermi Gamma-ray Burst Monitor found a weak transient signal~\citep{GW150914_GBM}, GW150914-GBM, though no corresponding signal was observed in other gamma-ray instruments \citep{Hurley:2016,Savchenko:2016, Tavani:2016}, nor was a similar signal found in relation to the other unambiguous GW detection in O1, GW151226~\citep{GW151226,Adriani:2016,Racusin:2017}, though $85\%$ of the LIGO localization for GW151226 was out of view by \Fermi, occulted by the Earth \citep{Racusin:2017}. 
%The relative sensitivity of the different gamma-ray instruments to GW150914-GBM-like events is still under investigation, and $85\%$ of the LIGO localization for GW151226 was out of view by \Fermi, occulted by the Earth \citep{Racusin:2017}.

The case for a GW candidate too low in significance to be claimed as an unambiguous detection would be strengthened by an EM detection consistent with the putative GW event. To that end, this paper reports on the search of GBM data for potential EM counterparts to any GW binary merger candidates found in the O1 data set with false alarm rate (FAR) below $10^{-5}$ Hz by the LIGO-Virgo analyses. Two GW candidates were identified in coincidence with gamma-ray signals exceeding the GBM background: a gamma ray transient found in the follow up of GW150914, previously reported in \cite{GW150914_GBM}, and a second GBM transient most likely due to a chance coincidence with a solar flare. 
%The search algorithms which produced the results reported here will continue to be refined and employed during GW observing campaigns.  
%During LIGO's second science run (O2) the GBM follow-up searches were be performed in near real time (latencies less then $\mathcal O(100)$ min) to rapidly identify high-energy EM emission from GW sources.

The paper is organized as follows.  In Section \ref{methods} we describe the data analysis methods used to identify compact binary coalescence (CBC) triggers in GW data and to identify gamma-ray transients in the GBM data.  Section \ref{results} presents the search results.  Section \ref{conclusions} summarizes the results and conclusions, and discusses the role for GBM and LIGO-Virgo joint analysis during future observing runs.

\section{Data Analysis Methods}\label{methods}

\subsection{Advanced LIGO}
The LIGO observatories are two modified Michelson interferometers located in Livingston, Louisiana and Hanford, Washington.  Each arm of the interferometer is formed by two mirrors separated by 4 km.  Passing GWs produce a strain, typically denoted as $h$, which changes the separation of the mirrors, inducing a relative phase difference in the light when it returns to the beam-splitter which transmits an optical signal proportional to the incident strain.  The large separation between the observatories reduces the false-alarm background due to coincident instrumental or environmental contamination, and differences in time and phase of arrival constrain the sky location of astrophysical sources.  For a full description see \cite{aLIGO}.

The data are analyzed for continuous~\citep{S6_Pulsar,O1_EinsteinAtHome}, 
stochastic~\citep{O1_Stochastic},
and transient astrophysical sources.  
Transient searches include ``all sky'' searches for new GW sources, and ``targeted'' searches for known astrophysical objects, e.g. GRBs~\citep{O1_GRB} and supernovae~\citep{S6_SNe}.  
All sky sources can be divided into two broad classes:  
Transients for which the gravitational-wave signal is well modeled, enabling the use of template waveforms in matched filtering searches for binary black holes~\citep{O1_BBH}, binary neutron stars and neutron star--black hole systems~\citep{O1_NSBH}, and cosmic strings~\citep{S6_CosmicString}; and ``unmodeled'' sources for which excess power searches are used~\citep{O1_Burst}.  

This work focuses on follow-up searches in GBM data of results from the template-based all sky search for CBC events, i.e. transient signals from the final stages of binaries containing compact stellar remnants (BHs and/or NSs) as gravitational-wave emission drives the objects to coalescence.  Because the GWs from CBC sources are well modeled and computationally tractable (see for example \cite{Blanchet:2006zz, Buonanno:1998gg, Ajith:2007qp}), data analysis strategies for such sources rely on template waveforms.

The candidate sources for GBM follow-up were acquired from two independently developed CBC search pipelines, \PyCBC \citep{DalCanton:2014, Usman:2016, Nitz:2017} and \GstLAL \citep{Privitera:2014, Messick:2017}. Here we briefly summarize both searches. For a more detailed  descriptions see \cite{GW150914_CBC} and \cite{O1_BBH}. Both \PyCBC and \GstLAL use accurate models of the gravitational waveform from coalescing binaries to perform a phase-coherent matched filtering search. The mass and spin parameter space of plausible CBC signals is covered by a grid of $\mathcal{O}(10^5)$ points and a \emph{template waveform} is computed at each point. 
The pipelines cover a total mass range between 2\msun and 100\msun with mass ratio from 1 to ${\sim}100$. The component spins are assumed to be parallel to the orbital angular momentum. 
Their dimensionless magnitudes can range from 0 to ${\sim}0.99$ for components heavier than 2\msun and from 0 to 0.05 for components lighter than 2\msun.  
Each template is correlated with the detectors' strain data to calculate the signal-to-noise ratio (\SNR). Data for which the same template produces a \SNR above the pre-established detection threshold in each detector within $15$ ms (slightly longer than the light travel time between detectors), are promoted to \emph{coincident triggers}  and ranked with a ``network statistic'' which, while different for \PyCBC and \GstLAL, is a function of the single-detector {\SNR}s and a measure of how consistent the data are with the template waveform.  Both pipelines estimate the background distribution of the network statistic in the absence of GWs, which is then used to map the network rank of each coincident trigger to the FAR of the search.  In the remainder of this work, the FAR from the GW searches will be referred to by \FARGW.  Candidates with \FARGW less than one per month exceeded the threshold for alerting EM observing partners participating in the O1 follow-up  program (e.g. ~\cite{GW150914_EM}).  \Fermi GBM has a unique data-sharing agreement with the LIGO Scientific and Virgo Collaborations, and has access to all CBC coincident triggers.

Because of their different assumptions and implementation details, \PyCBC and \GstLAL may produce different sets of triggers from the same data. Strong CBC signals are ranked by either pipeline to correspond to sufficiently low \FARGW, although not numerically equal between pipelines. However, a weak signal can in principle be detected by only one pipeline. For this reason, here we combine the triggers produced by both pipelines into a superset of CBC candidates. For duplicate triggers found by both pipelines, we select the candidate with the lowest \FARGW. The time range covered by the trigger set is from September 12, 2015 to January 19, 2016, with a total two-detector live-time of ${\sim}50$ days. Approximately $10^4$ triggers were recorded by \PyCBC and \GstLAL. In keeping with the spirit of an ``eyes-wide-open'' search for gamma-ray counterparts, no selection cuts were made to select for CBC candidates favoring systems theorized to produce potential electromagnetic counterparts (i.e., binaries containing at least one NS and, in the case of NSBHs, a sufficiently comparable mass ratio to tidally disrupt the NS).  

The CBC searches were repeated when improvements to the calibration of the detectors or the configuration of the pipelines were available. The triggers employed in the joint analyses were obtained from the first analysis using the offline CBC pipelines, and an intermediate calibration of the data.  Data produced with the final calibration were not available at the onset of this study. 
%We verified that there were no significant differences in the list of candidates used in this paper and that which was produced using the final calibration. 
%Triggers from the final runs of \PyCBC and \GstLAL, produced using the final calibration of the detectors and with \FARGW estimated using the entirety of the O1 data, were available with a much larger latency and are not used here. 
The candidate list used here was compared to that produced using the final calibration, and the only differences between the two were consistent with random noise fluctuations around the detection threshold of the CBC searches.
%\note{Tito--Can you say something about the overlap between trigger lists between different calibrations?}
%\FARGW corresponding to the ranking statistic of each trigger was calculated using a 15-day subset of the data around the trigger. 

%In preparation for the joint analysis with GBM, each CBC trigger in the sample was analyzed to produce a 2D probability density map the source sky location using the \BAYESTAR algorithm \citeme{cite:Singer:2015ema}. \BAYESTAR is method for producing low-latency, statistically robust, posterior probability maps for the sky location of CBC triggers by \note{using the template that produced the CBC trigger and analytically marginalizing over the time, phase, and amplitude over a finely sampled grid on the celestial sphere}.  \BAYESTAR sky-maps have larger statistical error than the maps produced by the Bayesian stochastic sampling follow-up analyses which marginalize over the full CBC parameter space~\citep{LALInference} but at a small fraction of the latter's computational cost.  \BAYESTAR maps are available within tens of seconds to minutes of a trigger, and are the natural choice for low-latency online EM follow-up and/or mass-production of example GW sky-maps. The final sky-maps can take hours to days to produce.  

%Not discussed here are searches for gravitational-wave counterparts to detected GRBs during O1, the results of which are reported in \cite{O1_GRB}.

\subsection{\Fermi GBM} 
The \Fermi Gamma-Ray Burst Monitor is an all-sky instrument onboard NASA's \Fermi Gamma-ray Space Telescope.  The GBM has an instantaneous field of view that subtends 70\% of the celestial sphere (with the remaining 30\% occulted by the Earth), with an 85\% live-time, and energy coverage from 8 keV to 40 MeV.  These capabilities make GBM an ideal partner in the searches for joint detections with ground-based GW observatories. The omnidirectional, broad energy coverage is accomplished through the use of 12 sodium iodide (NaI) detectors and two bismuth germanate (BGO) detectors.  

The responses for the NaI detectors are strongly dependent on the incident angle to the source, while the responses for the BGO detectors are not. Additionally, responses from all detectors are dependent on the source spectrum, since the blockage or attenuation of photons by the spacecraft and scattering off spacecraft components are energy dependent. These properties of the response will cause a signal to appear with a different strength in each detector, and the relative observed strength between all detectors allows the localization of signals with an accuracy on the order of a few to tens of degrees \citep{Connaughton:2015}. In practice, the expected photon rates at each detector are calculated over a grid of possible source spectrum templates and sky locations, and the observed relative rates are then compared to this expectation.

There are currently three distinct searches used on GBM data in the joint analysis of GW sources:   The onboard \emph{triggering} algorithm for automatic detection of strong GRBs~\citep{GBM,vonKienlin:2014,Bhat:2016};  the ground-based \emph{untargeted} search for weak SGRBs; and the \emph{targeted} search used to follow-up GW candidates~\citep{Blackburn:2015}. The onboard triggering and ground-based untargeted GBM searches are the means by which SGRBs are identified in the GBM data without any input about when or where a possible GW source may have occurred.  The targeted search of GBM data uses as input the arrival time of a GW candidate and uses the angular and energy dependencies of the detector response to find spectrally- and spatially-coherent weak signals.

\subsubsection{Blind, Untargeted GRB searches}
%The GBM flight software will trigger on-board when count rates in time-binned data from two or more NaI detectors exceed the background count rate based on the average of the previous ${\sim}15$ s of data. 
The GBM flight software will trigger on board when count rates in the time-binned data of two or more NaI detectors exceed the background count rates, determined by averages of $\sim{15}$ s of the previous data, by pre-specified thresholds (typically 4.5 sigma).
%The triggering threshold is bin-duration (16 ms to 4.096 s) and energy (50 - 300 keV, $>$100 keV, $>$300 keV) dependent. 
The on-board trigger is activated by approximately 40 SGRBs/year. For a full description of GBM and the on-board triggering algorithm see \citep{GBM,vonKienlin:2014,Bhat:2016}.

The untargeted, ground-based analysis searches primarily for SGRBs in continuous time-tagged event (CTTE) data.
%, which has been downlinked since November 2012 \note{Is the preceding sentence about CTTE data important?}. 
While a full description of the untargeted search will be in a forthcoming article we present a brief description here. The primary science goal of the search is to rapidly identify SGRBs missed by the triggering algorithm in order to maximize the potential for multimessenger observations, either through GW follow-up searches or to identify counterparts to astrophysical neutrinos.  The analysis searches for signals with durations ranging from 64 ms to 32 s. %, using four steps in phase.
%\note{``phase'' might be a bit jargon-y.  Is it important enough to warrant another sentence describing what is meant, or should we leave it out?}. 
Candidates are identified when two NaI detectors have an excess number of counts, with one exceeding 2.5 standard deviations ($\sigma$) relative to the background, and at least one other exceeding 1.25 $\sigma$. Correcting for the number of time bins in a day, the Poisson probability of the excess in the two detectors must be less than $10^{-6}$.
%There is additionally a requirement of the Poisson probability of the excess, corrected for the number of trials in a day.} 
%The search is biased towards events with favorable geometry to two detectors.  
The untargeted search is currently finding approximately 80 SGRB candidates per year in addition to those found by the on-board triggering algorithm\footnote{\url{https://gcn.gsfc.nasa.gov/admin/fermi_gbm_subthreshold_announce.txt}}.

\subsubsection{Follow-up Targeted GRB search}
The targeted search was developed to examine GBM data for gamma-ray transients in coincidence with GW candidates from LIGO's sixth science run (S6) and the concurrent second and third science runs (VSR2 and VSR3, respectively) of the Virgo 3 km interferometer in Cascina, Italy \citep{Virgo}. A description of how this method was applied to the S6+VSR2/3 data is presented in \cite{Blackburn:2015}. 

The targeted search identifies candidate gamma-ray transients over a 60 s window centered on the GW trigger time.  The 60 s time window is a conservative choice to not exclude precursor EM emission from the merger, or a delay between the compact objects' coalescence and launching of the relativistic jet.  Operating during O1, the targeted search utilized the daily continuous time (CTIME) data, with a nominal time resolution of 256 ms and 8 energy channels. The search was performed on binned data with bin-widths ranging from 256 ms to 8.192 s. 
%For CBC events there is presumably low circumburst densities, and the destruction of the neutron star should occur close in time to the merger, suggesting the time offset between the prompt SGRB emission and the merger time should be small. However, there is evidence for precursor emission in SGRBs which may be do to the fracturing of the neutron star crust some seconds before merger. Some SGRBs may also have extended emission that may be less beamed than the initial spike. The 30 second range should encompass all of these possibilities, with some buffer for unexpected results. 

%For the triggered GBM search we use the \BAYESTAR because tas we needed thousands of skymaps generated, and because the \BAYESTAR maps are generated quickly enough to be used in online searches for EM counterparts, allowing us to use them in future LIGO runs.

%The targeted search uses three template spectra for the analysis. 
The targeted search uses three template spectra to coherently forward model the detector response.
The templates are Band spectra \citep{Band:1993} comprised of two power-law components with spectral indices $\alpha$ and $\beta$ smoothly joined at the energy, $E_{\rm peak}$.  The definition of $E_{\rm peak}$, and therefore the spectral model, are modified from the original definition in \cite{Band:1993} and are explicitly shown in \cite{Connaughton:2015}.  The templates are referred to as ``soft,'' ``normal,'' and ``hard'' spectra with Band function parameters $\alpha$, $\beta$, $E_{\rm peak}$ = (-1.9, -3.7,70 keV), (-1, -2.3, 230 keV), and (0, -1.5, 1 MeV) respectively.  Each of the templates are utilized independently during the search, and then the most significant spectrum is identified for each time bin of the search.

The detection statistic $\Lambda$ for the targeted search is derived from a log likelihood ratio formalism that measures the GBM signal strength relative to a polynomial fit to the background count rate.  A detailed discussion of the search's methodology can be found in \cite{Blackburn:2015}. 

%The search uses three template spectra, soft, normal, and hard, to cover the range of possible spectra of GRBs. These have different distributions of significance so they are treated separately. We do not attempt to have a representative template bank for GRB spectra as there is no theoretical predictive basis to work from. As GBM is also background dominated we assign significance using the FAR, here referred to as FAR$_{GBM}$ for clarity, with each template spectra having a distinct FAR$_{GBM}$ distribution.

\begin{figure}
	\includegraphics[angle=270,width=0.5\textwidth]{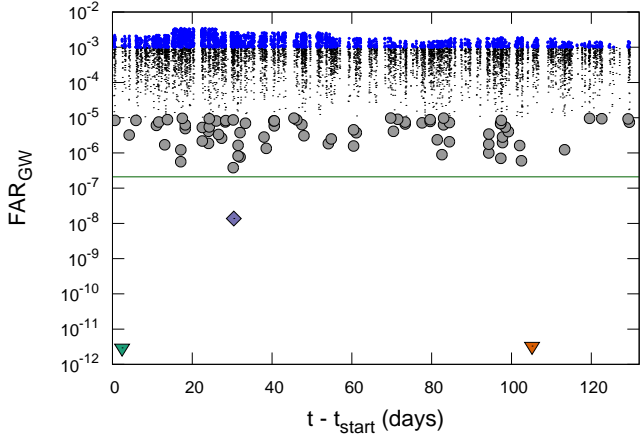}
	\caption{Time series of CBC triggers relative to the start-time of the joint analysis (September 12,windows 2015). The horizontal line denotes the inverse live-time for the search sample (${\sim}50$ days). The joint GBM/LIGO analyses used the blue points as the background sample, with \FARGW $>10^{-3}$ Hz. The gray circles correspond to the  search sample, with \FARGW $<10^{-5}$ Hz, or approximately fewer than one per day. The green and orange downward-pointing triangles correspond to the upper limits on \FARGW for GW150914 and GW151226, respectively. The purple diamond corresponds to LVT151012.  These colors and marker style are preserved throughout the article.}
	\label{fig:LIGO_long}
\end{figure}

\section{Search Results} \label{results}
\subsection{Advanced LIGO Sample}
Figure \ref{fig:LIGO_long} shows times relative to the start of the search sample (September 12, 2015) for the combined CBC triggers produced by the \GstLAL and \PyCBC pipelines. Intervals with no triggers correspond to times when one or both detectors were unable to produce science data, or when the data were vetoed due to well-understood instrumental or environmental disturbances. With the exception of the two high-confidence detections GW150914 (green triangle) \citep{GW150914} and GW151226 (orange triangle) \citep{GW151226}, and the low-significance candidate LVT151012 (purple diamond) \citep{O1_BBH}, no CBC triggers were more significant than the inverse live-time of the search ${\sim}2 \times 10^{-7}$ Hz (black horizontal line).  For the targeted GBM search, the 81 subthreshold candidates with \FARGW $<10^{-5}$ Hz ($\lesssim$ 1 day$^{-1}$) marked by gray circles were used as the search sample, while the 2935 blue dots (\FARGW$>10^{-3}$ Hz) served as the background sample. The candidates between the background and search samples (black dots) were discarded for two reasons.  First, to avoid the scenario where the background and search samples were sufficiently close in significance that the determination of which sample a candidate belonged in was up to random chance.  Second, due to the long tails in the GBM search statistics, the threshold used for the background sample provided sufficient triggers to measure the significance of anything in the search sample, making the inclusion of additional background triggers unnecessary.
\subsection{Triggering and Untargeted Search}

The simplest step in the joint analysis is to compare the list of times of known gamma-ray transients to the list of times of CBC candidates and search for a match. This comparison is complementary to the dedicated follow-up of known GRBs in GW data of GRBs, as reported in ~\cite{O1_GRB} in two ways: first, here we do not impose a maximum time offset between the GW and gamma-ray candidates, and second, this study includes sub-threshold GBM candidates from the untargeted search which, in O1, were not incorporated in the LIGO-Virgo GRB follow-up searches.

For the joint analysis, events found in the GBM triggering and untargeted searches during O1 are combined and correlated with the combined list of CBC candidates, searching for coincident events in the two observatories.  The GBM trigger list includes 115 GRBs, about half of which are SGRBs.  The search statistic is the shortest absolute time offset between a GRB and the search sample of CBC triggers, allowing for the possibility that the EM signal can precede or follow the compact binary merger event.

%The three closest coincidences came from the GBM untargeted search,  which recorded SGRBs 5, 11, and 28 s prior to a CBC candidate. All three of the CBC candidates had \FARGW $\gtrsim10^{-4}$ Hz. The triggered GBM event closest in time to a CBC trigger was GRB 151222340 which occurred ${\sim}20$ seconds before a 
%\GstLAL 
%CBC trigger with \FARGW $>10^{-4}$ Hz. 
%There are no GRBs in our search sample that occur shortly after a CBC trigger \note{define ``shortly''}.
%Due to the large time offset and the high \FARGW this is likely a random coincidence.

%0.00053564657268         -20.3600000739          464454178.550
%0.000305754905825        -4.8599999547   470745948.950
%0.000791043603265        -19.0839999318          ['GRB151222340' '472464619.016']

\begin{figure}
	\includegraphics[angle=270,width=0.5\textwidth]{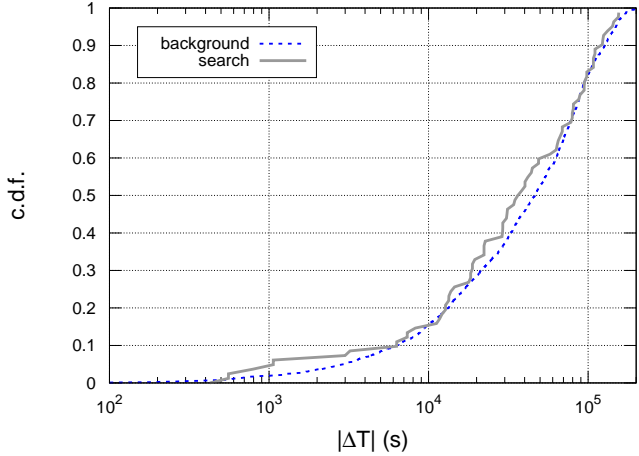}
	\caption{Cumulative distribution function of minimum absolute offset time between a GRB found by the triggered or untargeted GBM searches and a CBC candidate. The search sample (gray solid line) is consistent with the background distribution (blue dotted line), indicating that there are no significant coincidences between independently identified GBM and CBC triggers during O1.}
	\label{fig:timeoff}
\end{figure}

%The expected offset distribution is difficult to predict, owing to the long breaks in LIGO livetime \note{I don't see how the breaks in the LIGO livetime matter.  Assuming no correlation, time from CBC trigger to nearest GRB should be something like a nearest-neighbor distribution of a Poisson process over the LIGO livetime, no?  Can we have a PDF of $|\Delta t|$?  Maybe we can compare that to the null-hypothesis?  Overkill?? (Probably)} 

Using CBC candidates with \FARGW $>10^{-3}$ Hz as a background sample (blue dots in Fig. \ref{fig:LIGO_long}) and \FARGW $<10^{-5}$ as the search sample, we find the $90^{\rm th}$ percentile of the absolute offset time from the closest GRB in our sample to be ${\sim}10^4$ s. Approximately 10\% of the search sample are closer in time to a GRB, as expected for a null result based on the background sample.  The agreement between cumulative distribution functions of the absolute time offset $|\Delta T|$ for the background sample (blue dotted line) and the search sample (gray solid line) is shown in Figure \ref{fig:timeoff}. The shortest interval between the event times of a triggered GRB and a CBC candidate was $|\Delta T|=426.26$ s.

%As there are no promising signals we did not require location information. In the future a putative associated event would require agreement in localizations between instruments.

\subsection{Targeted Search}

\begin{figure}
        \includegraphics[angle=270,width=0.5\textwidth]{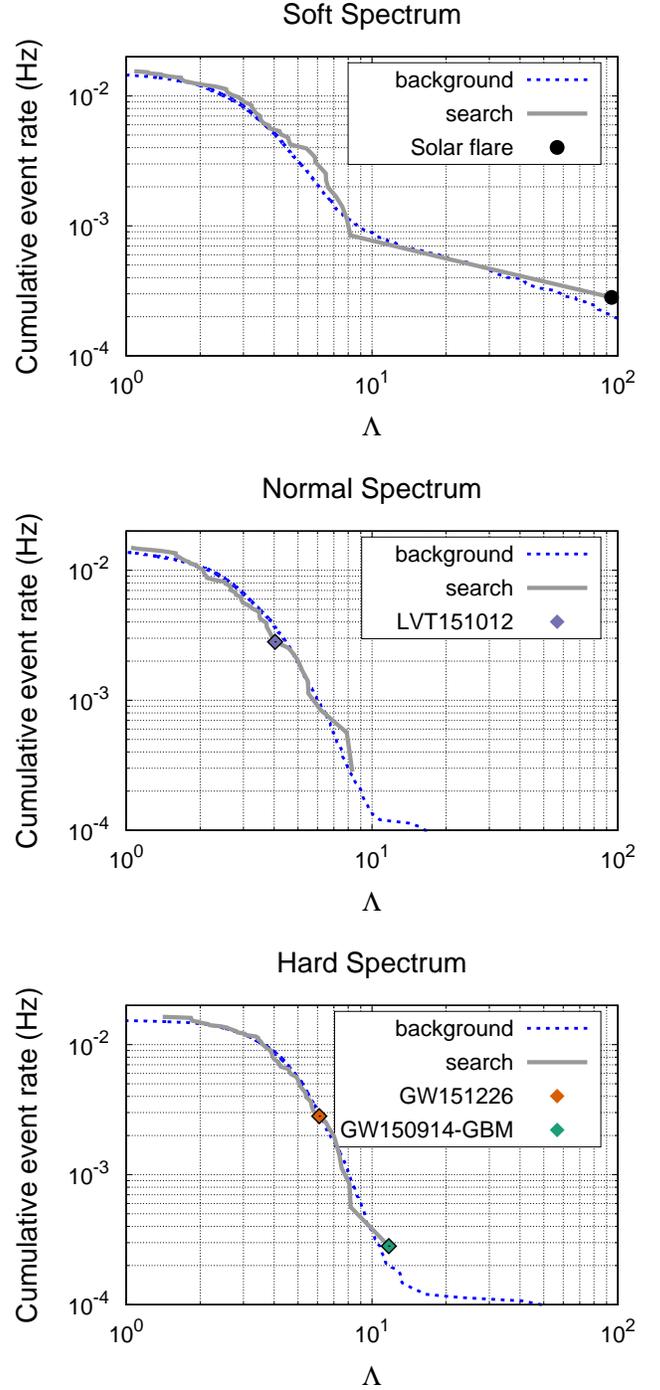}
	\caption{Cumulative event rates of the background (blue dotted lines) and search (gray solid lines) samples for the GBM targeted follow-up of CBC triggers as a function of the detection statistic $\Lambda$.  The panels in the figure from top to bottom correspond to results from the soft, normal, and hard template spectra. The distribution of the search samples is largely consistent with that of the background.  The two most significant events from the search distribution are the GBM transient identified in the follow-up to GW150914 (green diamond) and a soft, long-duration, transient found 26 s after a low-significance GW trigger.  The long, soft transient is likely due to solar activity, and in a chance coincidence with a GW trigger.}
	\label{fig:gbm_cumulative}
\end{figure}

The targeted search follow-up of \PyCBC and \GstLAL CBC coincident triggers used the same selection criteria as described earlier, with an additional down-selection of triggers which appeared in both CBC pipelines within 1 s of each other, in which case the trigger with the highest \FARGW was removed from the sample.  By this procedure, there were 17 times that a trigger was found by both pipelines:  Nine triggers with no offset ($|\Delta T|<0.01$ s), seven triggers with offset of $|\Delta T| \sim 0.01$ s, and one occurrence with an offset of $|\Delta T| \sim 0.02$ s. Of the remaining times of interest, 10\% occur when \Fermi was not taking data due to transit of the South Atlantic Anomaly (SAA), consistent with expectations. There were no cases where \Fermi entered or exited the SAA during the 60 s search interval. For the remaining GW triggers with GBM data, GBM observed between 52\% and 92\% of the GW skymap probability, with an average observing fraction of 69\%. Figure \ref{fig:gbm_cumulative} shows the cumulative event rate of the GBM background and search samples as a function of the detection statistic $\Lambda$. The distribution of the search samples is largely consistent with that of the background.
%Triggers with \FARGW $>10^{-3}$ Hz were used as the background sample to measure the false alarm rate for the GBM follow-up, \FARGBM.

%\begin{figure*}
%        \includegraphics[angle=270,width=\textwidth]%{targeted_search_logLR.eps}
	%\includegraphics[width=\textwidth, angle=270]{plots/targeted_search.eps}
%	\caption{Cumulative event rates of the background (blue dotted lines) and search (gray solid lines) samples for the GBM targeted follow-up of CBC triggers.  The panels in the figure from left to right correspond to results from the soft, normal, and hard template spectra. The distribution of the search samples is largely consistent with that of the background.  The two most significant events from the search distribution are the GBM transient identified in the follow-up to GW150914 (green diamond) and a soft, long-duration, transient found 26 s after a low-significance GW trigger.  The long, soft transient is likely due to solar activity, and in a chance coincidence with a GW trigger.}
%	\label{fig:gbm_cumulative}
%\end{figure*}

The GBM FAR, \FARGBM, for coincidences in the targeted follow-up of CBC candidates with \FARGW$<10^{-5}$ Hz can be mapped to a false alarm probability \FAP. Here the \FAP\ is similar to the quantity defined as the false alarm probability in \cite{GW150914_GBM}, which used an analytic approximation to estimate the significance of a candidate.  

However, the significance calculation used in this work differs from that in \cite{GW150914_GBM}.  For a given candidate found in the GBM follow-up search, the False Alarm Probability (FAP) is intended to report the probability of detecting a false associated signal (either statistical fluctuation or real unassociated signal) with a ranking that is at least as significant as the candidate.  

As in \cite{GW150914_GBM}, search candidates are first ranked by $R=1/({\rm FAR}_{\rm GBM} \vert\Delta T\vert)$ where \FARGBM\ is the cumulative event rate of the background distribution at the candidate's value for the search statistic (see Fig. 3), and $\vert \Delta T\vert $ is the absolute value of the time offset between the closest GW trigger time and the GBM candidate.  We impose a minimum $\Delta T_{\rm min}$ equal to the bin-width of the GBM data, which is 256 ms. 
%Thus we assume that the more significant a signal is by itself the more more likely it is to be an unambiguous detection, and the nearer in time the GBM candidate is to a GW trigger the more likely the association.  

For this analysis we empirically measure the FAP by using background noise events from the GW searches to seed GBM follow-up. Background events are ranked using the same statistic $R$.
The FAP of a candidate event is the $p$-value of the background distribution with that candidate's ranking statistic, $R$.

\begin{figure}
	\includegraphics[angle=270,width=0.5\textwidth]{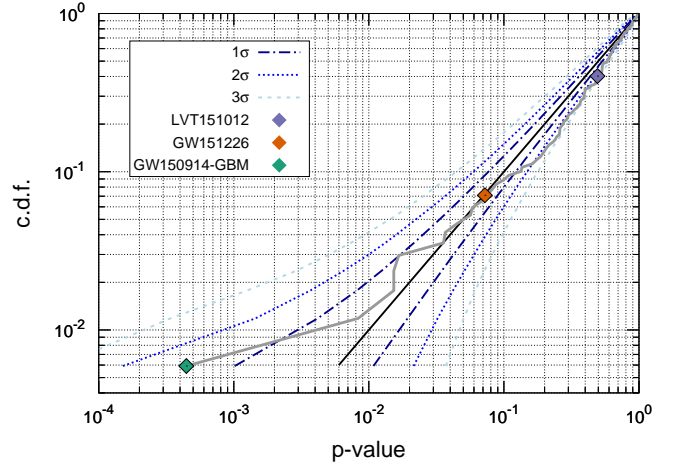}
	\caption{Cumulative distribution function of empirically-determined of \FAPs\ for the targeted search.  The black solid line corresponds to the null hypothesis that the search sample is consistent with background. The blue dashed lines envelop the $1\text{-}$, $2\text{-}$, and $3\text{-}\sigma$ confidence intervals. The highest ranking (lowest \FAP) event is GW150914-GBM (green diamond) which, for this search, has a significance of ${\gtrsim}1.5\sigma$.}
	\label{fig:targeted_FAR_FAP}
\end{figure}

Figure \ref{fig:targeted_FAR_FAP} shows the cumulative distribution of \FAPs\ for all of the CBC triggers follow-up candidates. GW150914-GBM had the lowest \FAP\ of any follow-up event, due to the fact that it had both the lowest \FARGBM\ and had the shortest time offset to its GW trigger time.
The \FAP\ for GW150914-GBM is an upper bound, because although there were more significant events in the background sample as determined by detection statistic $\Lambda$, the short time offset from the GW trigger caused it to have a higher ranking statistic, $R$, than any of the events in the background distribution. The blue dashed lines envelop the expected $1\text{-}$ (dark blue) $2\text{-}$ (blue) and $3\text{-}\sigma$ (light blue) confidence intervals assuming that the search distribution is statistically identical to the background distributions based on the number of events analyzed. 
From this it is clear that GW150914-GBM is a ${\gtrsim}1.5\sigma$ event. 
This is a lower significance than reported in \citet{GW150914_GBM} owing to the large number of CBC events in the search sample,  due to the low threshold chosen for trigger selection. This was intentional, as the point of this study was to search for sub-threshold joint-detections and we treat all CBC candidates in the search sample as equally worthwhile of follow-up.

%\begin{figure}
%	%\includegraphics[width=15cm]{paper_plots2/Seeded_search_FAR_FAP.pdf}
%	\includegraphics[angle=270,width=0.5\textwidth]{plots/FAP.eps}
%	\caption{Scatter plot of the combined search sample's \FARGBM vs.  absolute time offset between the CBC trigger and the GBM transient $|\Delta T|$.  The green diamond is the GBM transient detected in the follow-up of GW150914, \new{which is the most significant GBM event by an order of magnitude, though the significance is not sufficient to unambiguously claim it as a counterpart to the GW signal}.}
%	\label{fig:targeted_FAR_FAP}
%\end{figure}

%\begin{figure}
%	%\includegraphics[width=15cm]{paper_plots2/Seeded_search_FAR_FAP.pdf}
%	\includegraphics[angle=270,width=0.5\textwidth]{fap/FAPvT.eps}
%	\caption{Scatter plot of the combined search sample's FAP vs. absolute time offset between the CBC trigger and the GBM transient $|\Delta T|$, colored by \FARGBM.  The triangle is the upper limit for the GBM transient detected in the follow-up of GW150914, which is the most significant GBM event by \newtoo{at least} an order of magnitude, though the significance is not sufficient to unambiguously claim it as a counterpart to the GW signal.}
%	\label{fig:targeted_FAR_FAP}
%\end{figure}

%The lowest \FARGBM of the search came from the GBM transient found in the follow-up to the CBC trigger associated with GW150914.  The ranking statistic of the GBM source, GW150914-GBM, was measured to occur in the search with a \FARGBM of $2.0\times10^{-4}$ Hz and \FAP$=5\times10^{-1}$.  \FAP{} for the follow-up to GW150914 is higher in this analysis than  the false alarm probability reported in \cite{GW150914_GBM} because of the trials factor of $\mathcal{O}(100)$ due to the size of the search sample.
   
%Five other CBC candidates yielded coincidences in the GBM triggered search with \FARGBM$<10^{-3}$ Hz. 
Also note that the second lowest false alarm rate (\FARGBM=$2.6\times10^{-4}$ Hz) of the search occurred in the analysis of the GBM transient detected at 2015-09-29 12:15:43.6 UTC.  The event was at the edge of the search window, 26 s before the GW trigger time that prompted the GBM analysis, resulting in \FAP${\sim}1$. The event was found using the soft template and the longest (8.192 s) timescale of the search.  A search for other possibly associated astrophysical transients at that time revealed that the GBM event occurred during the exponential decay tail of an M-class solar flare (e.g. GBM triggers 150929346 and 150929553).  The GBM localization for this candidate is consistent with the Sun and inconsistent with the LIGO sky map, thus we conclude this GBM event is due to solar activity.  

\section{Conclusions and Discussion} \label{conclusions}
%In this paper we have reported on results from the search of \Fermi GBM data for electromagnetic counterparts to GW candidates. We performed two separate searches for gamma-ray signals in the \Fermi GBM around triggers from the \GstLAL and \PyCBC searches for compact binary merger signals in LIGO data collected during O1:  A blind search correlating the catalog of GBM GRBs to subthreshold GW triggers; and a targeted coherent GBM follow-up seeded by the LIGO event times.   Results from both searches were largely consistent with background estimated by using high \FARGW candidates.  No on-board triggered GRBs, or GRBs found by the untargeted GBM search, were found to be within ${\sim}100$ s of a CBC trigger.  The most significant event in the targeted follow-up was from the trigger associated with GW150914.  The GBM source, GW150914-GBM, was previously reported in \cite{GW150914_GBM}.  The event's false alarm rate reported here, \FARGBM $=2.2\times10^{-4}$ Hz, was determined from an empirical measurement of the search background over the entire O1 run, and was found to be of comparable significance to the previous estimate of $1.2\times10^{-4}$ Hz using only the 3 day period around the GW detection.  No other CBC candidates were found to have notable counterparts in the GBM follow-up search.

The GBM follow-up analyses searched for gamma-ray transients associated with a list of CBC candidates during O1.  These CBC candidates were well below the standards for the GW trigger alone to be considered likely due to a compact merger, but significant enough that if a gamma-ray transient was found by GBM in coincidence, it would support an astrophysical origin of the GW transient.
%any CBC candidates which were above a significance threshold well below the standards for the GW trigger alone to be considered likely due to a compact merger. Had a significant gamma-ray transient been found by GBM in coincidence with a GW candidate, either through the independent triggering and untargeted searches, or the targeted follow-up of GW candidates, it would have provided evidence for an astrophysical origin of the GW transient.

There were no coincidences between GBM transients found by the triggered or untargeted searches and the LIGO CBC candidates. For the targeted follow-up of GW candidates, GW150914-GBM is the most significant event of the search. The analysis presented here was not designed to revisit the significance of the GW150914-GBM association; additional observations of BBH mergers with GBM will be needed to establish or rule out the astrophysical nature of GW150914-GBM. No other significant transients were found in the GBM targeted search, ruling out the possibility of a low-significance gravitational-wave event being identified by its association with a gamma-ray transient during O1.

The significance of GW150914-GBM reported by the search described in this paper (${\gtrsim}1.5\sigma$) is conservative in part because of the choice to not include in the candidate ranking the significance of the GW triggers.
%themselves, instead treating each CBC candidate--despite many having very low significance from the CBC searches--as being equally likely to produce a gamma-ray transient. 
By not using the GW significance in our assessment of the validity of a GBM counterpart, the targeted search is hampered by implicitly placing marginal GW events on equal footing with more plausible ones.  Joint GW/EM events may prove to be rare, and the search strategy as adopted is to cast a wide, model-independent, net when looking for possible sub-threshold coincidences in both observatories. Deriving a joint significance which includes information about the likelihood of the GW candidate being astrophysical is subtle; having a high-significance GW event does not necessarily imply a high likelihood for an associated gamma-ray transient.  Incorporating the significance of the GW candidate into the GBM follow-up analysis is an area of investigation for future joint observing campaigns~\citep{Ashton:2017}.

For the targeted search, the long tails in the background distribution out to high values of the ranking statistic limited the sensitivity of the search in O1.  Further analysis has revealed that many ($>20\%$) of the high likelihood events in the background can be attributed to large background fluctuations due to high particle activity associated with \Fermi's proximity to the SAA.  For analyses of later observing runs, the background fitting procedure and the vetoing of data from entry and exit of the SAA have been improved~\citep{Goldstein:2016}.

The multimessenger observations of the binary neutron star merger associated with a short GRB, and the resulting kilonova and GRB afterglow observations, demonstrated a best-case scenario of a bright, nearby transient which was identified unambiguously gravitationally and electromagnetically. However, the event had the closest recorded redshift for any GRB despite its relatively low brightness~\citep{GRB170817A} while the binary neutron star system was well within the detection horizon for the LIGO-Virgo network~\citep{GW170817_BNS}. 
It is therefore not difficult to imagine scenarios where sub-threshold searches would have been critical for identifying the association.
Joint analyses that can elevate the significance of candidate events are needed to best exploit the rich datasets from gravitational-wave and gamma-ray survey instruments. The searches reported here, while not producing any unambiguous counterparts, will serve as a foundation for future joint analysis.

\acknowledgments%
The USRA co-authors gratefully acknowledge NASA funding through contract NNM13AA43C. The UAH co-authors gratefully acknowledge NASA funding from co-operative agreement NNM11AA01A. EB and TDC are supported by an appointment to the NASA Postdoctoral Program at the Goddard Space Flight Center, administered by Universities Space Research Association under contract with NASA. DK, CAWH, CMH, and TL gratefully acknowledge NASA funding through the \Fermi GBM project. Support for the German contribution to GBM was provided by the Bundesministerium f{\"u}r Bildung und Forschung (BMBF) via the Deutsches Zentrum f{\"u}r Luft und Raumfahrt (DLR) under contract number 50 QV 0301. AvK was supported by the Bundesministeriums f{\"u}r Wirtschaft und Technologie (BMWi) through DLR grant 50 OG 1101. NC and JB acknowledge support from NSF under grant PHY-1505373. SMB acknowledges support from Science Foundation Ireland under grant 12/IP/1288.

The authors gratefully acknowledge the support of the United States
National Science Foundation (NSF) for the construction and operation of the
LIGO Laboratory and Advanced LIGO as well as the Science and Technology Facilities Council (STFC) of the
United Kingdom, the Max-Planck-Society (MPS), and the State of
Niedersachsen/Germany for support of the construction of Advanced LIGO 
and construction and operation of the GEO600 detector. 
Additional support for Advanced LIGO was provided by the Australian Research Council.
The authors gratefully acknowledge the Italian Istituto Nazionale di Fisica Nucleare (INFN),  
the French Centre National de la Recherche Scientifique (CNRS) and
the Foundation for Fundamental Research on Matter supported by the Netherlands Organisation for Scientific Research, 
for the construction and operation of the Virgo detector
and the creation and support  of the EGO consortium. 
The authors also gratefully acknowledge research support from these agencies as well as by 
the Council of Scientific and Industrial Research of India, 
the Department of Science and Technology, India,
the Science \& Engineering Research Board (SERB), India,
the Ministry of Human Resource Development, India,
the Spanish  Agencia Estatal de Investigaci\'on,
the  Vicepresid\`encia i Conselleria d'Innovaci\'o, Recerca i Turisme and the Conselleria d'Educaci\'o i Universitat del Govern de les Illes Balears,
the Conselleria d'Educaci\'o, Investigaci\'o, Cultura i Esport de la Generalitat Valenciana,
the National Science Centre of Poland,
the Swiss National Science Foundation (SNSF),
the Russian Foundation for Basic Research, 
the Russian Science Foundation,
the European Commission,
the European Regional Development Funds (ERDF),
the Royal Society, 
the Scottish Funding Council, 
the Scottish Universities Physics Alliance, 
the Hungarian Scientific Research Fund (OTKA),
the Lyon Institute of Origins (LIO),
the National Research, Development and Innovation Office Hungary (NKFI), 
the National Research Foundation of Korea,
Industry Canada and the Province of Ontario through the Ministry of Economic Development and Innovation, 
the Natural Science and Engineering Research Council Canada,
the Canadian Institute for Advanced Research,
the Brazilian Ministry of Science, Technology, Innovations, and Communications,
the International Center for Theoretical Physics South American Institute for Fundamental Research (ICTP-SAIFR), 
the Research Grants Council of Hong Kong,
the National Natural Science Foundation of China (NSFC),
the Leverhulme Trust, 
the Research Corporation, 
the Ministry of Science and Technology (MOST), Taiwan
and
the Kavli Foundation.
The authors gratefully acknowledge the support of the NSF, STFC, MPS, INFN, CNRS and the
State of Niedersachsen/Germany for provision of computational resources.

\bibliographystyle{aasjournal}
\bibliography{/Applications/TeX/papers}
%\printbibliography

\end{document}